\documentclass[runningheads]{llncs}
\usepackage[T1]{fontenc}

\usepackage{graphicx}
\usepackage{url}
\usepackage{graphicx}
\usepackage{caption}
\usepackage{subcaption}
\usepackage{multirow}
\usepackage{float}
\usepackage{hyperref}
\usepackage{amsfonts}
\usepackage{amsmath}
\usepackage{lscape}
\usepackage{xcolor}

\begin{document}
\title{
AttackER: Towards Enhancing Cyber-Attack Attribution with a Named Entity Recognition Dataset
}

\titlerunning{AttackER: Towards Enhancing Cyber-Attack Attribution with NER}

\author{Pritam Deka, Sampath Rajapaksha
\and
Ruby Rani
\and Amirah Almutairi
\and \\
Erisa Karafili%\inst{1}%\orcidID{2222--3333-4444-5555}
}
\authorrunning{ } 
\institute{University of Southampton\\ University Road, Southampton SO17 1BJ, UK \\
\email{\{p.deka, srwg1m24, r.rani, a.almutairi, e.karafili\}@soton.ac.uk}}
\maketitle

\begin{abstract}
Cyber-attack attribution is an important process that allows experts to put in place attacker-oriented countermeasures and legal actions. The analysts mainly perform attribution manually, given the complex nature of this task. AI and, more specifically, Natural Language Processing (NLP) techniques can be leveraged to support cybersecurity analysts during the attribution process. However powerful these techniques are, they need to deal with the lack of datasets in the attack attribution domain. In this work, we will fill this gap and will provide, to the best of our knowledge, the first dataset on cyber-attack attribution. We designed our dataset with the primary goal of extracting attack attribution information from cybersecurity texts, utilizing named entity recognition (NER) methodologies from the field of NLP. Unlike other cybersecurity NER datasets, ours offers a rich set of annotations with contextual details,  including some that span phrases and sentences. We conducted extensive experiments and applied NLP techniques to demonstrate the dataset's effectiveness for attack attribution. These experiments highlight the potential of Large Language Models (LLMs) capabilities to improve the NER tasks in cybersecurity datasets for cyber-attack attribution.

\end{abstract}

\keywords{Attribution, Dataset, NLP, Named Entity Recognition, LLMs}

\section{Introduction}
Attribution can be defined as ``determining the identity or location of an attacker or an attacker’s intermediary'' \cite{wheeler2003techniques}. This non-trivial process is crucial for responding to incidents, formulating cybersecurity policies, and addressing future threats. Knowing who the attacker is allows the defender to put in place attacker-oriented countermeasures and when possible legal actions. 
Cyber-attack attribution is a complex and highly resource-consuming process \cite{perry2019no} as several measures can be taken by threat actors to hide their traces. Furthermore, it is often difficult to identify threat actors as they may share common targets and tools to carry out attacks, resulting in a limited degree of confidence in the attribution output. 

 Cybersecurity experts perform the attribution process manually, as currently there is no tool that can automate or provide support for such a complex process. Another critical issue related to cyber-attack attribution is the complexity and massive amount of information that analysts need to look online to be updated with the new trends of attacks, their analyses, and past attributions \cite{rid2015attributing}. 
Security experts rely on AI and ML techniques and tools to identify, filter, and cluster the evidence left after an attack \cite{sarker2021ai}. However, these solutions are limited for the cyber-attack attribution problem as there are currently no datasets that incorporate data that can be useful for this process.

The advent of deep neural architectures such as transformers 
have made it possible to achieve state-of-the-art results in various tasks of Natural Language Processing (NLP) such as text classification, text summarization, and named entity recognition (NER)~\cite{devlin2018bert}. Although NER is one of the fundamental tasks of NLP, 
it is underexplored in domains like cybersecurity due to the complex nature of the involved entities \cite{gasmi2018lstm}. 
NER is valuable in identifying various entities from text sources which can then be analysed for further information. Current NLP techniques can be leveraged to support the attribution process by automatically identifying entities involved in the cyber-attack. However, there is currently no dataset for cyber-attack attribution that can be used for training NLP models to automatically identify significant information to help and automate this complex and time-consuming process.

In this work, we will address the above problems by constructing the first cyber-attack attribution dataset, called \emph{AttackER}, using NER. AttackER will fill the research gap in the domain of cyber-attack attribution. Our dataset will allow the automatic identification of various entities based on AI and NLP methods, thereby providing support to the experts and reducing the time and resources for the attribution process. Our dataset can be used in the future to train models that automatically extract useful insights from human/machine generated text and reports that can be later fed to an automatic tool for cyber-attack attribution and investigation. Furthermore, these insights can be used by security analysts to be up-to-date with new cyber-attacks and threats. Thus, drastically reducing the resources and the time needed to perform the attribution process. AttackER is the first step in using AI and NLP techniques to support and in the future automate the attribution process.

In this paper, we also provide the analyses performed on the dataset and the models trained on it. For our analyses, we used transformer-based model training like Huggingface (HF) and spaCy, as well as fine-tuned LLMs like GPT-3.5, Llama-2, and Mistral-7B, for the NER task. The positive results of our analyses (especially LLMs), highlight the effectiveness of AttackER.

\textit{\textbf{Our contributions}}: Our main contribution in this paper is AttackER\footnote{\url{https://zenodo.org/records/10276922}}, a new dataset for entity recognition in attack attribution and investigation tasks. This dataset, which consists of 18 distinct types of entities, helps identify a wide range of objects in cybersecurity text. Thus, it assists in providing a wealth of information for detailed analysis and research. Along with the dataset we highlight below the other contributions of this paper:
\begin{itemize}
    \item Enhance the performance of LLMs on the NER task of the AttackER dataset through instruction fine-tuning using specific prompt templates. To the best of our knowledge, this is the first study to explore the use of fine-tuned LLMs for NER in cyber-attack attribution.
    \item 
    We are also releasing Huggingface transformer-based models trained on AttackER, along with the dataset itself, to further research in this domain. Both the models and the dataset are publicly available\footnote{\url{https://huggingface.co/Cyber-ThreaD}}.
\end{itemize}
In Section~\ref{sec:relatedwork}, we provide the related work in this research domain. We introduce our dataset and the entities used in Section~\ref{sec:dataset}. We show the methodology used for the annotation and the model training in Section~\ref{sec:methodology}. In Section~\ref{sec:experiments}, we introduce and discuss our experimental results. 
We finally conclude and discuss future works in Section~\ref{sec:conclusion}.

\section{Related Work}\label{sec:relatedwork}

NER in cybersecurity has been extensively studied, primarily focusing on cyber threat intelligence. Early work, like \cite{mulwad2011extracting}, introduced a framework using SVM to extract vulnerability and attack information from web text. 
Similarly, \cite{joshi2013extracting} proposed an approach to identify entities, concepts, and relationships associated with cybersecurity within the text using Conditional Random Field (CRF). 
These approaches relied on hand-crafted features, requiring significant feature engineering effort. 
With the rise of neural networks, studies like \cite{gasmi2018lstm} and \cite{gasmi2019information} explored using LSTM-CRF architectures by blending LSTM, word2vec embeddings~\cite{mikolov2013distributed}, and CRF for cybersecurity NER and demonstrated their effectiveness compared to traditional methods. 
With the advent of transformer \cite{vaswani2017attention} neural network architecture and its advantages over other neural network architectures, much of the research on NER in the domain of cybersecurity has shifted towards the usage of models based on transformer neural network architecture. The authors of \cite{evangelatos2021named} explored the usage of transformer models in NER task of cyber threat intelligence text. 
In~\cite{tikhomirov2020using}, the authors use BERT-based models for NER on the Russian cybersecurity domain. They compared three different models where two are pre-trained over Russian text. Of these two, one of the models was pre-trained specifically on cybersecurity text. Through experimental results, the authors showed that the model pre-trained specifically on cybersecurity text performed the best out of the three models. 

Recently, there has been a significant advancement in the field of LLMs, and it has shown excellence in a variety of NLP tasks. So far, minimal research has been conducted in the direction of NER using LLMs. Wang et al.~\cite{wang2023gpt} proposed GPT-NER, a task generation NER to fill this gap, achieving competitive performance with supervised models, particularly in low-resource environments. Refined NER performs well on benchmarks but poorly on unknown entities. Zhang et al.~\cite{zhang2024linkner} presented LinkNER, which combines optimised models with LLMs and an uncertainty-based approach that yields better performance, especially when managing noisy web data pertinent to cybersecurity. The application of LLMs to NER in cybersecurity faces challenges due to a task mismatch. Recent works by \cite{wursch2023llm} and \cite{shafee2024evaluation} employed LLMs such as ChatGPT for the NER task in cybersecurity, highlighting the limitations of LLMs for cybersecurity data and demonstrating their unreliability for cybersecurity NER. To the best of our knowledge, no existing work has applied fine-tuned LLMs for NER in the cybersecurity domain, particularly for cyber-attack attribution. In this paper, we address this gap by fine-tuning three LLMs including GPT-3.5, Llama-2, and Mistral-7B on our created AttackER dataset.

Publicly available datasets such as DNRTI \cite{wang2020dnrti}, APTNER \cite{wang2022aptner}, and CyNER \cite{alam2022cyner} have been introduced for training NER models in cybersecurity. However, none of these datasets are developed with a focus on cyber-attack attribution and investigation. In contrast, the AttackER dataset is specifically designed using NER to capture information about attack attribution.

\section{Dataset details}\label{sec:dataset} 
Let us now introduce the AttackER dataset for attack attribution and investigation. Unlike existing datasets, AttackER offers highly detailed information through a more complex annotation process, providing context beyond words or tokens regarding their respective entity types. This allows state-of-the-art transformer models to learn more robustly and provide deeper insights for cybersecurity analysts.
To the best of our knowledge, no other datasets have been produced for the NER task in this domain, focusing on the analysis of cyber-attacks and their attribution. %Let us now explain in detail how we created the dataset. 

To define the entity types of our dataset, we used the STIX 2.1 framework\footnote{\url{https://oasis-open.github.io/cti-documentation/stix/intro}}. 
 The STIX 2.1 objects represent different elements of cyber threats, such as indicators, threat actors, campaigns, tools, etc. which help in organizing and categorizing threat information consistently. 
Although STIX 2.1 defines 18 different objects, we did not include all of them, as some of these objects were not useful and were never used during the cyber-attack attribution process, as not pertinent. Thus, we include only 14 STIX 2.1 objects. Based on the Ontology for cyber-attack attribution introduced in~\cite{dilpret}, we added two new objects for the identification of entities that are not part of the STIX 2.1 but are important for the cyber-attack attribution process. The added entities deal with information like the impact of the attack (``IMPACT''), and the motivations (``ATTACK\_MOTIVATION'').

During our exploratory analysis of the objects, we found that two of the used STIX 2.1 objects could be ambiguous for the purpose of attack attribution. Thus, we divided them (``Tool'' and ``Identity'') into sub-classes. We divided ``Tools'' into ``GENERAL\_TOOLS'' and ``ATTACK\_TOOLS'' to distinguish between general tools and tools used for attacks. For the object ``Identity'', we divided it into ``VICTIM\_IDENTITY'' and ``GENERAL\_IDENTITY'' which separate the victims of an attack from other identities. We created this level of granularity for ``Tools'' and ``Identity'', as it is required during the attribution process. This division was made to remove ambiguity and specify the entities more robustly, thereby providing more nuanced information. In total, we have a set of 18 entities for the task of NER in cyber-attack attribution. Unlike other datasets in the domain of cybersecurity for NER, we have a rich set of annotations with contextual details and some of the annotations span across phrases and even sentences. This is particularly useful when we want information like the impact of an attack or the course of action taken after an attack. 
Our dataset covers 2640 sentences annotated with the entity labels out of these 18 different entity types. 

\subsection{Entity set description}
\begin{figure}[h]
    \centering
    \includegraphics[width=0.7\textwidth]{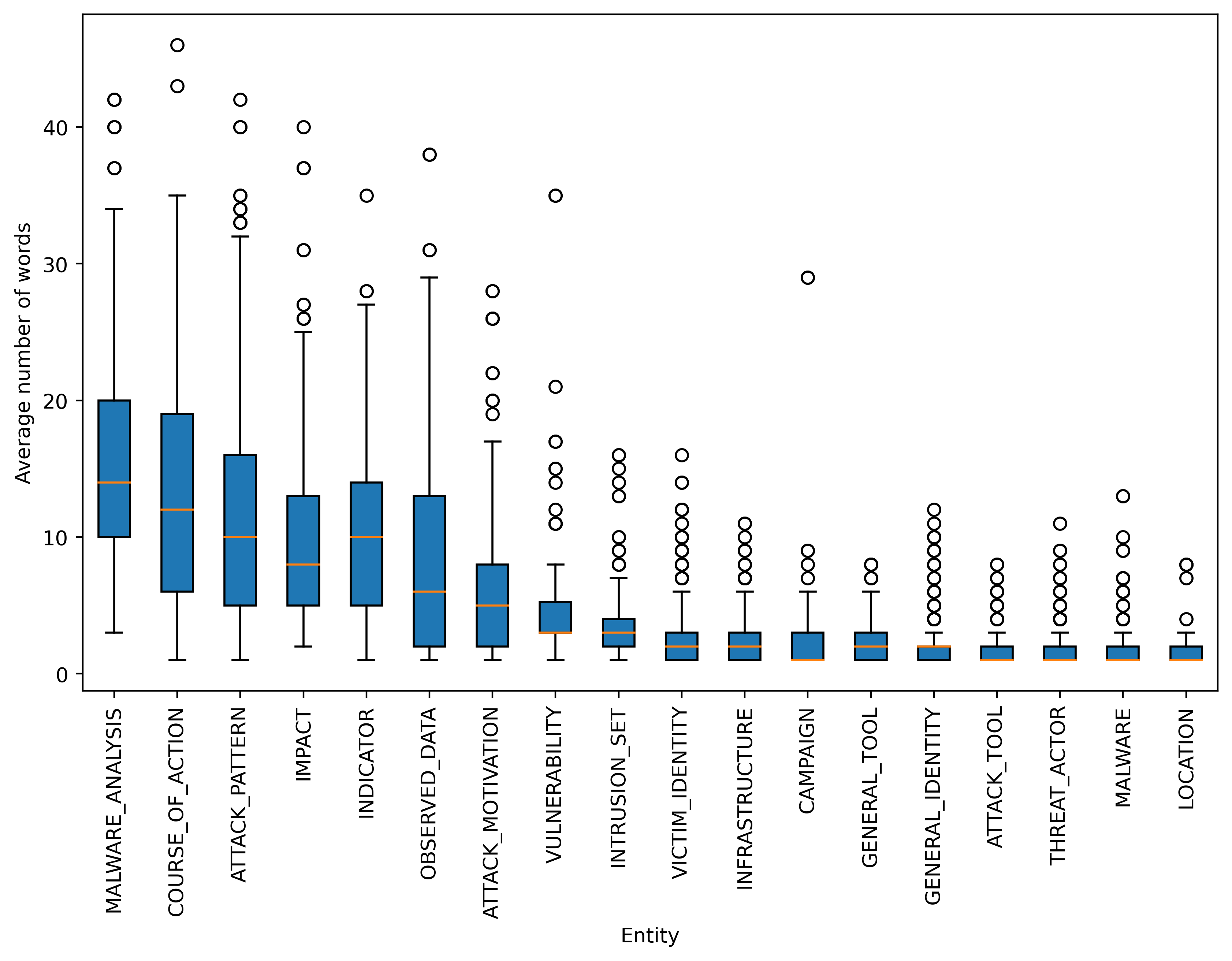}
    \caption{Average length of the entities}
    \label{entity}
\end{figure}

We have a set of 18 entities to label 
text extracted from various reports and blogs that deal with cyber-attack attribution and investigation. 
We followed manual labelling to assign pertinent entities to each text and later on continued with a semi-automatic approach (see  Section \ref{spacy}).
The details of the entities used for our dataset creation are presented in Table \ref{dataset_details_table}. Distribution for the average number of words for each entity type is shown in Figure \ref{entity}. Notably, entity types such as ``MALWARE\_ANALYSIS'' and ``COURSE\_OF\_ACTION'' encompass multiple words, while others like ``MALWARE'' and ``LOCATIONS'' consist of fewer words. By utilising these 18 labels, we 
annotated texts through reading and understanding, assigning each text the relevant label.
\begin{table}[t]
\centering
\begin{tabular}{|l|l|}
\hline
\textbf{Entity Type}               & \textbf{Count} \\ \hline
ATTACK\_PATTERN               & 948            \\
GENERAL\_IDENTITY             & 879            \\
INFRASTRUCTURE                & 473            \\
INDICATOR                     & 152            \\
GENERAL\_TOOL                 & 250            \\
COURSE\_OF\_ACTION            & 325            \\
THREAT\_ACTOR                 & 814            \\
VULNERABILITY                 & 236            \\
MALWARE\_ANALYSIS             & 157            \\
INTRUSION\_SET                & 165            \\
VICTIM\_IDENTITY              & 458            \\
MALWARE                       & 648            \\
IMPACT                        & 181            \\
ATTACK\_TOOL                  & 271            \\
OBSERVED\_DATA                & 368            \\
LOCATION                      & 235            \\
CAMPAIGN                      & 244            \\
ATTACK\_MOTIVATION            & 222            \\
\hline
\textbf{Total count of entities}       & 7026           \\ \hline
\end{tabular}
\caption{Entity details for the AttackER dataset}
\label{dataset_details_table}
\end{table}

We describe below the used entities that are important pieces that help us understand and analyse cyber threats: 
\begin{itemize}
 \item ATTACK\_PATTERN: Shows how attackers use techniques, tactics, and procedures to compromise targets.
 \item GENERAL\_IDENTITY: Gives information about identities not directly involved in cyber-attacks, providing context.
  \item CAMPAIGN: Describes a group of harmful activities or attacks happening over a certain time, having specific targets. 
 \item COURSE\_OF\_ACTION: Includes plans to stop or respond to attacks, using technical measures and broader actions like training or policy changes.
 \item VICTIM\_IDENTITY: Identifies entities that got compromised during cyber-attacks or targeted by the threat actor.
 \item INDICATOR: Contains patterns used to find suspicious or harmful cyber activity, like sets of bad domains.
 \item  INFRASTRUCTURE: Represents systems, software services, and resources supporting different purposes, including C2 servers and targeted devices. 
 \item INTRUSION\_SET: Groups together harmful behaviors and resources believed to be organised by one organisation.
 \item LOCATION: Shows geographic places relevant to cyber threats. 
 \item MALWARE: Represents malicious code.
 \item MALWARE\_ANALYSIS: Captures details and results from checking malicious code, whether by looking at its code or watching it run. 
 \item ATTACK\_TOOL: Software used by malicious actors for attacks and tools for running campaigns.
 \item GENERAL\_TOOL: Identifies tools on systems used for normal uses, e.g., remote access, checking networks.
 \item VULNERABILITY: Points out weaknesses or problems in software or hardware that can be used by attackers.
 \item IMPACT: Includes possible effects of cyber threats e.g., physical/digital, economic, psychological, social/societal aspects.
 \item  ATTACK\_MOTIVATION: Provides the motivations behind an attack. 
\item OBSERVED\_DATA: Gives information related to cybersecurity, e.g., file names, systems, and networks.
\item THREAT\_ACTOR: Represents real people, groups, or organisations believed to be acting with malicious intentions.
\end{itemize}

\section{Methodology}\label{sec:methodology}
In this section, we discuss the data annotation process and NLP models, including spaCy, HF transformers, and the LLMs fine-tuning procedure for our dataset. The three different classifications represent the progression and diversity of NLP technologies. 

\subsection{Annotation of data}
To create AttackER, we automatically collected documents from various online sources. These documents pertain to reports, articles, and blogs about previous attacks written by cybersecurity experts. In these documents, past cyber-attacks were analysed and when possible attributed. We selected only documents that an analyst would be interested in reading and gathering information from (overall 217 documents). We performed a manual analysis of the collected documents where we checked that the content was suitable and made sure that they were analysing different types of attacks.  
The major sources from where we collected these documents include blogs and reports from Mandiant, Malwarebytes, Mitre, Securelist, Trendmicro. A full list is provided in Appendix \ref{appendix}.

We used scraping text tools, like two different Python libraries %in order to scrape text from these websites, 
namely newspaper3k\footnote{\url{https://github.com/codelucas/newspaper}} and BeautifulSoup\footnote{\url{https://www.crummy.com/software/BeautifulSoup/}}, to collect the data. We then performed text pre-processing on the scraped text to remove unwanted information like blank spaces, unwanted characters, and text from images and advertisements on the websites. We did not remove any stopwords or punctuation marks, as they were needed for the annotation phase. Automatic pre-processing was not applicable to all the scraped text as there were several web sources and it became difficult to tailor automatic pre-processing techniques for all of them. Thus, manual inspection was also carried out along with the automatic techniques. 
For the annotation part, we used a tool called Prodigy\footnote{\url{https://prodi.gy/}} 
which provides a feature-rich annotation interface and a seamless integration with the spaCy \cite{honnibal2020spacy} library. 

For the annotation process, we had two human annotators doing the labelling. Given the complexity and high technicality of the analysed text, we decided to have two cybersecurity experts 
conducting the annotations. 
We calculated an agreement score between the annotators and had a couple of rounds where the annotation was checked by more senior cybersecurity experts. Previous research~\cite{yim2016tumor} 
has shown that, for NER tasks, instead of agreement scores such as Cohen's Kappa, 
methods like partial matches are more relevant due to the complex nature of the task. We calculated the partial match F-1 score between the two annotators for 321 sentences and found an average F-1 score of 85.83\% between the two annotators showing a high partial match agreement score. Upon the establishment of an agreement between the annotators, the annotation was carried out. Once we annotated a reasonable representation for each entity, 
we used the spaCy library to train a transformer model on the annotated data which is then used for a semi-automatic annotation approach. The model annotated data were all reviewed and when needed corrected by our human annotators.

After compiling the annotated data and creating our dataset, we experimented with spaCy library, HF transformers 
(HF) library and the fine-tuned LLMs which includes GPT-3.5, Llama-2, and Mistral-7B for NER. 
The motivation to perform these experiments was to analyse our dataset 
and how models trained on it will help analysts to analyse information related to cyber-attack attribution automatically. 

\subsection{spaCy model training}\label{spacy}
We explore various transformer models that encompass both general models and cybersecurity domain-specific models. 
We followed the spaCy training guidelines 
to train the transformer models over our annotated data for the NER task. 
The hyperparameters include a learning rate of 5e-05, a batch size of 128, a maximum sequence length of 128, and  
used the Adam optimizer. The loss function used during training is the categorical \emph{cross-entropy loss function}  which is used to measure the difference between predicted probabilities and the true distribution of the labels. 
We experimented with different splits and achieved the best results with the 
train-test split of 80:20. The 80\% of the data is used as training data and the 20\% is further split into a 70:30 ratio, where 70\% is used as the evaluation set and 30\% is used as the test set which is held out for prediction. 
After training is completed, the trained model is used to annotate new documents with the annotators' feedback. Once new documents are annotated, they are combined with the previous annotated documents, and the model is re-trained using the same process for an improved version.

\subsection{Huggingface model training}
We use the HF library 
as it provides an API based platform to access state-of-the-art open-sourced transformer models for NLP tasks, including NER. 
We changed the format of the data since the HF transformer library uses the IOB/BIO 
format. 
 We used the publicly available training script\footnote{\url{https://github.com/huggingface/transformers/blob/main/examples/pytorch/token-classification/run_ner.py}} from HF to train the models. We experimented with 10 epochs, a maximum sequence length of 512 and batch size 2 with a learning rate of 2e-05. For optimization, we used the Adam optimizer and a categorical cross-entropy loss function. We saved each checkpoint at 500 steps and used the best model from the saved checkpoints for prediction on the test set.

\subsection{Large language model fine-tuning}
Although autoregressive LLMs such as GPT-3 have demonstrated promising results in various NLP tasks, their performance in NER tasks still falls short of supervised baselines, primarily due to their focus on text generation over sequential labelling~\cite{wang2023gpt}. To address this, supervised fine-tuning can be employed to fine-tune a pre-trained model for specific tasks. In particular, we decided to experiment with LLMs to enhance the NER performance on AttackER by leveraging the reasoning capabilities of LLMs.
In our approach, we use a specialized fine-tuning technique called instruction fine-tuning to guide the LLMs' output towards the NER task requirements. In instruction-based fine-tuning, we provide prompts with examples to help models learn and improve response accuracy. For this purpose, the data format used in the HF experiments was transformed into an instruction dataset, which includes instructions along with the desired output conforming to the LLMs' accepted prompt template, as described in \cite{lyu2024keeping}.
We employed two open-source state-of-the-art LLMs for our NER task: Llama-2-7b-hf (Llama-2)\footnote{\url{https://huggingface.co/meta-llama/Llama-2-7b-hf}} and Mistral-7B\footnote{\url{https://mistral.ai/}} and one OpenAI LLM model i.e., GPT-3.5\footnote{\url{https://platform.openai.com/docs/models/gpt-3-5-turbo}}.
Given that LLMs outputs are influenced by the provided instructions, we experimented and selected the best prompt format for different LLMs. 
Table ~\ref{tbl:LLM_prompts} presents an example prompt template\footnote{All LLM models use the same prompt, with only syntactical differences for GPT-3.5.} for Llama-2~\cite{touvron2023llama}
and Mistral-7B~\cite{jiang2023mistral}
models.
Using the formatted dataset, we employed parameter-efficient fine-tuning coupled with quantized low-rank adaptation to enhance the efficiency and reduce the resource requirements of the fine-tuning process.
For the zero-shot learning experiments (these are the base models that use their existing knowledge and learn to solve unseen tasks) using the base Llama-2 LLM model, we provide the prompt template employed in Appendix~\ref{appendixcc}.

\begin{table}[h]
\centering
\scriptsize
\begin{tabular}{|l|p{11cm}|}
\hline
LLM&Prompt\\
\hline
Llama-2 & <s>[INST] You are a cybersecurity and NLP expert tasked with named entity extraction. You need to provide the named entities for each token in the given sentence. You will be penalized for incorrect predictions. Identify the entities for the following sentence: `This stolen credential access can also be used to launch a ransomware attack .' [/INST] [This:O, used:O, to:O, launch:O, a:O, ransomware:B-CAMPAIGN, attack:I-CAMPAIGN, .:O] </s> \\
\hline
\end{tabular}
\caption{LLMs prompt template} 
\label{tbl:LLM_prompts}
\end{table}

\section{Experimental Results and Discussion}\label{sec:experiments}
We now discuss the experimental results using spaCy, transformer-based models, and autoregressive LLMs for NER 
on our dataset. 
We evaluated various transformer models for NER, including generic models like RoBERTa%~\cite{liu2019roberta}
and DeBERTa-v3,%~\cite{he2021debertav3},
an enhanced version of DeBERTa%~\cite{he2020deberta}
while SecureBERT, %~\cite{aghaei2022securebert}, 
SecBERT, %~\cite{liberato2022secbert}, 
and CyBERT%~\cite{ranade2021cybert} 
are pre-trained on cybersecurity-specific text. Moreover, autoregressive LLMs models (Llama-2, Mistral-7B, and GPT-3.5) are exploited as generic models and further fine-tuned on AttackER dataset. We used the same train-test-eval split for all experiments. 

\subsection{spaCy and Huggingface experiments}

To evaluate the models, precision, recall, and F-1 score \cite{rainio2024evaluation} were used, with F-1 score as the final metric due to its harmonic mean property. A 100\% alignment between predicted and annotated tokens was required for entity-level accuracy assessment. The results for both spaCy and HF models are presented in Table~\ref{spacy_results} for the test dataset evaluation.

\begin{table}[t]
\centering
\begin{tabular}{|l|lll|lll|}
\hline
\multirow{2}{*}{\textbf{Models}} & \multicolumn{3}{l|}{\textbf{spaCy}}                                           & \multicolumn{3}{l|}{\textbf{Huggingface}}                                           \\ \cline{2-7} 
                                 & \multicolumn{1}{l|}{\textbf{P}} & \multicolumn{1}{l|}{\textbf{R}} & \textbf{F-1} & \multicolumn{1}{l|}{\textbf{P}} & \multicolumn{1}{l|}{\textbf{R}} & \textbf{F-1} \\ \hline
\textbf{SecureBERT}              & \multicolumn{1}{l|}{0.6647}     & \multicolumn{1}{l|}{0.6516}     & \textbf{0.6581}       & \multicolumn{1}{l|}{0.5194}     & \multicolumn{1}{l|}{0.5912}     & {0.5530}       \\ \hline
\textbf{SecBERT}                 & \multicolumn{1}{l|}{0.6689}     & \multicolumn{1}{l|}{0.5647}     & 0.6124       & \multicolumn{1}{l|}{0.4452}     & \multicolumn{1}{l|}{0.5735}     & 0.5013       \\ \hline
\textbf{CyBERT}                  & \multicolumn{1}{l|}{0.5809}     & \multicolumn{1}{l|}{0.4853}     & 0.5288       & \multicolumn{1}{l|}{0.2256}     & \multicolumn{1}{l|}{0.3882}     & 0.2854       \\ \hline
\textbf{RoBERTa}                 & \multicolumn{1}{l|}{0.6018}     & \multicolumn{1}{l|}{0.5735}     & 0.5873       & \multicolumn{1}{l|}{0.5485}     & \multicolumn{1}{l|}{0.6324}     & 0.5874       \\ \hline
\textbf{DeBERTa-v3}              & \multicolumn{1}{l|}{0.6489}     & \multicolumn{1}{l|}{0.6470}     & 0.6480       & \multicolumn{1}{l|}{0.5402}     & \multicolumn{1}{l|}{0.6324}     & 0.5827       \\ \hline
\end{tabular}%}
\caption{spaCy and Huggingface experimental results on the AttackER dataset}
\label{spacy_results}
\end{table}

Among the spaCy models tested, SecureBERT outperformed other BERT-based models, likely due to its training on a vast amount of cybersecurity data from various online sources, unlike other models trained on more specific cyber-related data. However, DeBERTa-v3, an improvement over BERT and RoBERTa, achieved a comparable F-1 score despite not being trained explicitly on cybersecurity data. Conversely, in the HF models, RoBERTa slightly outperformed DeBERTa-v3. Overall, the spaCy models demonstrated better performance than the HF models. Since default parameters were used for all models, it is possible that the spaCy models had more optimal hyperparameters for NER tasks compared to the HF models.

\subsection{LLMs Experiments}
For the LLMs, the base and fine-tuned models were employed to evaluate and compare the impact of fine-tuning on NER tasks for attack attribution. The base models are equivalent to their corresponding zero-shot learning~\cite{kojima2022large}. During data annotation, entities were assigned based on the most suitable match, where only one entity was selected for a word or phrase. However, a word or phrase might correspond to multiple potential entities. Our analysis of LLMs outputs revealed their ability to identify alternative valid entities distinct from the annotated ground truth. While these are classified as incorrect predictions compared to the annotated ground truth, they can also be considered accurate predictions. To reflect this, we manually reviewed all sentences in the test dataset and updated our ground truth entity types to match the LLMs' predictions when they represented another accurate entity type.
The experimental results of the LLMs on the test set are summarized in Table~\ref{LLM_results}. The adjusted ground truth results reflect the evaluation based on the revised test dataset, considering the accurate LLMs predictions.

\begin{table}[t]
\centering
\begin{tabular}{|l|lll|lll|}
\hline
\multirow{2}{*}{\textbf{LLM}} & \multicolumn{3}{l|}{\textbf{Ground Truth}}                                           & \multicolumn{3}{l|}{\textbf{Adjusted Ground Truth}}                                           \\ \cline{2-7} 
                                 & \multicolumn{1}{l|}{\textbf{P}} & \multicolumn{1}{l|}{\textbf{R}} & \textbf{F-1} & \multicolumn{1}{l|}{\textbf{P}} & \multicolumn{1}{l|}{\textbf{R}} & \textbf{F-1} \\ \hline
\textbf{Llama-2}              & \multicolumn{1}{l|}{0.5941}     & \multicolumn{1}{l|}{0.6013}     & \textbf{0.5976}       & \multicolumn{1}{l|}{0.7723}     & \multicolumn{1}{l|}{0.7509}     & \textbf{0.7615}       \\ 
\hline
\textbf{Mistral-7B}              & \multicolumn{1}{l|}{0.5601}     & \multicolumn{1}{l|}{0.5221}     & \textbf{0.5404}       & \multicolumn{1}{l|}{0.8670}     & \multicolumn{1}{l|}{0.6650}     & \textbf{0.7527}       \\
\hline
\textbf{GPT-3.5}              & \multicolumn{1}{l|}{0.6064}     & \multicolumn{1}{l|}{0.6099}     & \textbf{0.6081}       & \multicolumn{1}{l|}{0.8862}     & \multicolumn{1}{l|}{0.8172}     & \textbf{0.8503}       \\
\hline
\end{tabular}%}
\caption{LLMs experimental results on the AttackER dataset}
\label{LLM_results}
\end{table}

\subsubsection{Discussion of the results}
Llama-2 zero-shot learning attained an F-1 score of 0.4801, marking a 10\% decrease compared to the fine-tuned HF RoBERTa model. This can be attributed to the limited performance of LLMs in sequence labelling tasks. Instead, the fine-tuned model (see Table~\ref{LLM_results}) slightly surpassed the HF RoBERTa model. When evaluated using the adjusted ground truth dataset, there was a significant 28\% enhancement for the fine-tuned model compared to the zero-shot learning approach and a 17\% improvement relative to the HF RoBERTa model. The top-performing model on the AttackER dataset was the spaCy SecureBERT model. However, Llama-2 outperformed it with a 10\% improvement in the F-1 score. 
Mistral-7B's evaluation results show it underperforms on the ground truth on the AttackER dataset with an F-1 score of 0.5404. However, when evaluated using the adjusted ground truth, Mistral-7B shows a significant improvement while achieving an F-1 score of 0.7527. The evaluation results indicate that LLMs can identify multiple suitable tags for a token, and adjusting the ground truth significantly enhances the model's capabilities for the task. 
GPT-3.5 performed robustly, with an F-1 score of 0.6081 using the ground truth dataset. This performance is slightly better than the open-source LLMs' (Llama-2 and Mistral-7 B) ground truth evaluations. When assessed with the adjusted ground truth dataset, the F-1 score of 0.8503 was achieved, showing it as the top-performing model among the three. The remarkable outcomes of the model's high recall and precision values of 0.8172 and 0.8862, respectively, attest to its efficacy for NER tasks in the cybersecurity domain.

Tables~\ref{spacy_results} and \ref{LLM_results} demonstrate that the spaCy SecureBERT model achieved a noteworthy F-1 score of 0.6581, but much less than GPT-3.5's top performance when comparing the LLM results with the spaCy and HF models. The LLM models outperformed the HF models on adjusted ground truths, with RoBERTa and DeBERTa-v3 demonstrating competitive F-1 scores of 0.5874 and 0.5827, respectively.
Evaluation and comparison between HF, spaCy, and LLMs show that LLMs significantly differ from traditional BERT-based models in the spaCy and HF frameworks, especially GPT-3.5 and LLM models on adjusted ground truth. This suggests that LLMs can significantly enhance performance in specialised NER tasks for cybersecurity documents when adjusted and tested using adjusted ground truth. Improved precision and recall, coupled with the adaptive nature of LLMs, make these models an effective way to boost entity recognition accuracy in challenging and domain-specific datasets such as AttackER.

\begin{table}[t]
\centering
\resizebox{\textwidth}{!}{%
%\scriptsize
\begin{tabular}{|p{4cm}|p{4cm}|p{4cm}|}
\hline
Ground Truth& Llama-2 Predictions & Adjusted Ground Truth\\
\hline
We:O, observed:O, limited:O, instances:O, of:O, Russia:B-INTRUSION\_SET, affiliated:I-INTRUSION\_SET, domains:I-INTRUSION\_SET, being:O, targeted:O&
We:O, observed:O, limited:O, instances:O, of:O, Russia:B-VICTIM\_IDENTITY, affiliated:B-VICTIM\_IDENTITY, domains:B-VICTIM\_IDENTITY, being:O, targeted:O&
We:O, observed:O, limited:O, instances:O, of:O, Russia:B-VICTIM\_IDENTITY, affiliated:B-VICTIM\_IDENTITY, domains:B-VICTIM\_IDENTITY, being:O, targeted:O\\
\hline
\end{tabular}}
\caption{Capability of LLMs to identify accurately other entity types}
\label{tbl:Adjusted_GT}
\end{table}

In Table~\ref{tbl:Adjusted_GT} we show the annotated ground truth, Llama-2 predictions, and adjusted ground truth for a specific sentence. In this sentence, ``Russia affiliated domains'' could be categorized as either ``INTRUSION\_SET'' or ``VICTIM\_IDENTITY''. While entity annotation typically assigns only one entity per word or phrase, a word or phrase might correspond to multiple entities. LLMs can accurately identify entities that may not align with the original ground truth. Consequently, the ground truth is revised to match the LLMs’ precise predictions. For instance, the ground truth for ``Russia affiliated domains" is updated to ``VICTIM\_IDENTITY''. However, such modifications are made only when the LLMs predictions are accurate and diverge from the original annotations; otherwise, the original annotations are retained. Compared to LLMs, especially fine-tuned Llama-2, Mistral-7B, and GPT-3.5 predictions, none of the other models, including spaCy, HF transformer models, and base LLMs (zero-shot learning), produced accurate predictions for other entities to achieve a higher F-1 score with the adjusted ground truth.

\section{Conclusion}\label{sec:conclusion}
In this paper, we introduced a novel dataset about cyber-attack attribution, AttackER, using NER. To the best of our knowledge, this is the first dataset to include information that will be useful for the cyber-attack attribution process. 
We analysed our dataset and the models trained on it.
We presented three distinct approaches for our analyses on the NER task using spaCy, HF transformer-based models, and fine-tuning the LLM models including open-source LLMs; Llama-2 and Mistral-7B, and closed-source LLM OpenAI GPT-3.5. Our experiments revealed that GPT-3.5 zero-shot learning attained an F-1 score of 0.4546, marking a 13\% decrease compared to the fine-tuned HF RoBERTa model. When evaluated using the adjusted ground truth dataset, there was a significant 39\% enhancement for the fine-tuned model compared to the zero-shot learning approach and a 25\% improvement relative to the HF RoBERTa model. These results highlight the effectiveness of our dataset for the NER task when leveraging LLMs such as GPT-3.5, Llama-2, and Mistral-7B models. 
These results confirm that our novel dataset,
AttackER will help cybersecurity analysts during the attribution process using AI models trained on our contextually rich and informative dataset. 

In future works, we plan to increase the dataset size and focus on entities that have a lower count. We will explore different solutions for automating cyber-attack attribution using AttackER. Specifically, we plan to identify and extract useful relationships between the labelled entities that can provide meaningful insights to the analysts, thus, keeping them up-to-date and speeding up some of the tasks needed to perform the attribution process. Having models that automatically extract entities and relationships from existing reports/blogs can be used to enrich existing tools~\cite{karafili2020} that help cyber-attack attribution, but also to fully automate this complex process. The full automation of attack attribution will drastically reduce the time and resources needed. Furthermore, AttackER can be used for tools that support analysts during cyber-attack investigations and attribution. Thus, helping with resource cutting and working towards the reduction of bias and false attribution.  
Another interesting direction is to use our results to build a question and answering tool for cyber-attack investigation and attribution.

\section*{Acknowledgement}
Research funded by the University of Southampton on behalf of the Defence Science and Technology Laboratory (Dstl) which is an executive agency of the UK Ministry of Defence providing world class expertise and delivering cutting-edge science and technology for the benefit of the nation and allies. The research supports the Autonomous Resilient Cyber Defence (ARCD) project within the Dstl Cyber Defence Enhancement programme.
\bibliographystyle{plain}
\bibliography{AttackER}

\appendix
\section{Appendix A} \label{appendix}
Other sources from where we collected our data include BleepingComputer, HexaCorn, SentinelOne, CrowdStrike, Reuters, Att, Kaspersky, Webroot, Welivesecurity, Virusbulletin, Tadviser, Forumspb, Netresec, Brighttalk, Libevent, Fortinet, Microsoft, Washingtonpost, Reversemode, Viasat, Wikipedia, Wired, Cisa, Airforcemag, Businesswire, Cyberuk, Proofpoint, Fb, Withsecure, Techcrunch, Mozilla, Humansecurity, Nist, Intel471, Morphisec, Payplug, Sophos, Coretech, Stratixsystems, Crayondata, Medium, Cybergeeks, Gridinsoft, Securin, Rsisecurity, ITgovernance, DigitalGuardian, IronNet, ThreatConnect, ProtectUK, Forbes.

\section{Appendix B} \label{appendixcc}
Table~\ref{tbl:LLM_prompts0} shows an example of the prompt template utilized for zero-shot learning for Llama-2.

\begin{table}[h]
\centering
\scriptsize
\begin{tabular}{|l|p{11cm}|}
\hline
LLM&Prompt\\
\hline
Llama-2& You are an NLP expert tasked with CyberThreat-entity Entity Extraction. Identify entities of the types ATTACK\_PATTERN, GENERAL\_IDENTITY, INFRASTRUCTURE, INDICATOR, GENERAL\_TOOL, COURSE\_OF\_ACTION, THREAT\_ACTOR, VULNERABILITY, MALWARE\_ANALYSIS, INTRUSION\_SET, VICTIM\_IDENTITY, MALWARE, IMPACT, ATTACK\_TOOL, OBSERVED\_DATA, LOCATION, CAMPAIGN, and ATTACK\_MOTIVATION in the following input: 'Further analyses of these similarities are available via Mandiant Advantage.'
Your answer must be in the form of a [token 1:'tag', token 2:'tag', token 3:'tag']. If there's no entities for a token, assign 'O'. Take care; your answer is only valid if it follows the correct format! Provide the tag for each token in the sentence. Use 'B-' prefix for the starting word of an entity and 'I-' prefix for the words inside an entity. \\
\hline
\end{tabular}
\caption{LLM prompt template used for base (zero-shot learning) models}
\label{tbl:LLM_prompts0}
\end{table}
\end{document}